\documentclass[amsmath,nofootinbib,twocolumn]{revtex4-1}
\pdfoutput=1

\usepackage{aas_macros,graphicx,marvosym,subfigure}
\usepackage[colorlinks]{hyperref}

\newcommand{\br}[1]{\left[#1\right]}
\newcommand{\cu}[1]{\left\{#1\right\}}
\newcommand{\pa}[1]{\left(#1\right)}
\newcommand{\ed}{\,\mathrm{d}}
\newcommand{\pd}{\,\partial}

\begin{document}

\title{Near-horizon Kerr magnetosphere}

\author{Samuel E. Gralla$^\text{\Lightning\Heart}$}
\author{Alexandru Lupsasca$^\text{\Lightning}$}
\author{Andrew Strominger$^\text{\Lightning}$}
\affiliation{
\Lightning\,
Center for the Fundamental Laws of Nature, Harvard University, Cambridge, Massachusetts 02138, USA\\
\Heart\,
Department of Physics, University of Arizona, Tucson, Arizona 85721, USA
}

\begin{abstract}
	We exploit the near-horizon conformal symmetry of rapidly spinning black holes to determine universal properties of their magnetospheres.  Analytic expressions are derived for the limiting form of the magnetosphere in the near-horizon region.  The symmetry is shown to imply that the black hole Meissner effect holds for free Maxwell fields but is generically violated for force-free fields.  We further show that in the extremal limit, near-horizon plasma particles are infinitely boosted relative to accretion flow.  Active galactic nuclei powered by rapidly spinning black holes are therefore natural sites for high-energy particle collisions.
\end{abstract}

\maketitle

Emergent conformal symmetries near critical points play a profound role in modern condensed matter physics and particle physics.  A maximally spinning black hole can be viewed as a critical point of the Kerr family where the surface gravity and Hawking temperature vanish.  Reference~\cite{Bardeen:1999} identified a corresponding global conformal symmetry, in the sense that the enhanced near-horizon isometry group contains the conformal group $\mathsf{SO}(2,1)$.  The near-horizon isometries were later found in Ref.~\cite{Guica:2008} to be part of an even larger emergent infinite-dimensional local conformal symmetry.  This suggests that physics near rapidly spinning black holes should be constrained by conformal symmetry in much the same way as many near-critical condensed matter systems are governed by conformal field theories.  Indeed, the conformal symmetries have already proven to be powerful tools for near-horizon physics and astrophysics \cite{Porfyriadis:2014,Hadar:2014,Hadar:2015,Zhang:2014,Lupsasca:2014,Lupsasca:2015,Compere:2016}.

In this paper, we realize and explore a set of near-horizon symmetry constraints of potential astrophysical relevance.  We observe that the scaling used to define the near-horizon metric imposes scaling relations on any other physical fields present as well.  We show that tensor fields become self-similar in the limit, and that outside of special cases they also become \textit{null}, in the sense that all scalar invariants vanish.  We can then apply field equations on these simpler fields in the simpler near-horizon metric and learn nontrivial properties of the fields on the extreme Kerr background from which they descend.

As an example, we consider stationary, axisymmetric electromagnetic fields sourced by charge-current exterior to the extreme Kerr horizon.  The generic limiting field is incompatible with Maxwell's equations, implying that the limit must vanish.  A subleading limit vanishes as well, provided there is no net charge.  Back in terms of fields on extreme Kerr, these results imply that the tangential components of the Maxwell field vanish on the horizon.  This is a covariant formulation and generalization of the ``black hole Meissner effect'' \cite{Bicak:1985}, which states that maximally spinning black holes expel magnetic flux.  Thus, the Meissner effect may be viewed as a direct consequence of the conformal symmetry.

We next consider a black hole surrounded by a diffuse (force-free) plasma magnetosphere.  This configuration is generally believed to be the engine behind active galactic nuclei (AGN), since the plasma can efficiently extract energy from the black hole in steady state \cite{Blandford:1977}.  We then take a near-horizon limit and demand regularity on the future (but not the past) horizon.  The resulting force-free field configuration is parametrized by a free function of the polar angle $\theta$ and generically symmetric/self-similar under only two of the three $\mathsf{SO}(2,1)$ conformal symmetries of the near-horizon metric.  In stark contrast to the vacuum case discussed above, the near-horizon limit \textit{automatically} satisfies the force-free equations.  Thus, the symmetry imposes \textit{no} constraints on force-free fields in extreme Kerr.  This provides further evidence that there is no black hole Meissner effect for plasma \cite{Komissarov:2007} and that energy extraction can occur all the way to extremality.

The solution discussed above is singular on the past horizon, which is acceptable because astrophysical black holes do not have past horizons.  It is also interesting to consider what happens if we demand regularity on both the future and the past horizon.  This condition turns out to be very strong and completely fixes the $\theta$ dependence of the scaling solution, which then preserves all three of the $\mathsf{SO}(2,1)$ conformal symmetries.  Hence, every smooth solution of the force-free equations on extreme Kerr reduces (up to an overall constant) to the same form in the near-horizon limit.  The solution is electrically dominated $(F^2<0)$.  In flat Minkowski space, electrically dominated solutions are unstable to acceleration of charged particles.  The strong gravitational fields in the near-horizon region could potentially eliminate the instability.  Further work is required to settle the issue.

We also analyze the behavior of particle worldlines in the near-horizon limit.  Worldlines that plunge into the extreme Kerr horizon are infinitely boosted by the scaling and become null plunge trajectories in in the near-horizon metric, while worldlines that remain outside the extreme Kerr horizon have no near-horizon limit at all.  More interestingly, we may consider particles that are parametrically close to the extreme Kerr horizon, represented by a family of worldlines approaching the horizon generators.  Taking the near-horizon limit of such a family yields a timelike trajectory.  Intersections between these two classes of near-horizon trajectories describe the particle collisions of arbitrarily high-energy considered in Refs.~\cite{Piran:1975,Banados:2009}.  We may view the existence of such collisions as a consequence of the existence of a nontrivial near-horizon limit.

We conclude by generalizing some of these observations to near-extreme black holes and applying them to the Blandford-Znajek model of an AGN.  We point out that in the extremal limit, matter on the innermost stable circular orbit (ISCO) is at infinite relative boost to matter in the surrounding diffuse plasma.  This suggests that AGN with rapidly spinning black holes may naturally realize the high-energy collisions of Refs.~\cite{Piran:1975,Banados:2009}.

While we have focused on two cases (Maxwell fields and point particles) in this paper, we note that the technique is quite general.  All smooth tensor fields pick up additional symmetry in the limit, allowing simpler computations that can reveal universal properties.  The conformal symmetry provides an effective organizing principle for the near-horizon dynamics of rapidly spinning black holes.

In Sec.~\ref{sec:Limits}, we review the near-horizon limit of the extreme Kerr metric and explain how other fields behave under the limit.  In Sec.~\ref{sec:Maxwell}, we consider Maxwell fields and in Sec.~\ref{sec:Particles} we discuss high-energy particle collisions.  Finally, in Sec.~\ref{sec:AGN}, we generalize our discussion to include near-extreme black holes and note the possibility of high-energy collisions in AGN.  Our metric has signature $(-,+,+,+)$ and we use Heaviside-Lorentz units with $G=c=1$.

\section{Near-horizon limits}
\label{sec:Limits}

The Kerr metric for a black hole of mass $M$ and angular momentum $aM$ is given in Boyer-Lindquist (BL) coordinates $t,r,\theta,\phi$ by 
\begin{align}
\label{eq:Kerr}
	ds^2=&-\frac{\Delta}{\Sigma}\pa{\ed t-a\sin^2{\theta}\ed\phi}^2
		+\frac{\Sigma}{\Delta}\ed r^2+\Sigma\ed\theta^2\nonumber\\
		&+\frac{\sin^2{\theta}}{\Sigma}\br{\pa{r^2+a^2}\ed\phi-a\ed t}^2,
\end{align}
where $\Delta=r^2-2Mr+a^2$ and $\Sigma=r^2+a^2\cos^2{\theta}$.  Consider the extreme case $a=M$ and introduce the ``scaling coordinates"
\begin{align}
\label{eq:ScalingCoordinates}
	T=\frac{\lambda t}{2M},\quad
	R=\frac{r-M}{\lambda M},\quad
	\Phi=\phi-\frac{t}{2M}.
\end{align}
For small $\lambda$, these new coordinates cover only the region $r\to M$ near the horizon.  In the $\lambda\rightarrow0$ limit with $T,R,\theta,\Phi$ held fixed, one obtains the metric of the so-called Near-Horizon Extreme Kerr (NHEK) region,
\begin{align}
\label{eq:NHEK}
	ds^2=2M^2\Gamma\!\br{-R^2\ed T^2+\frac{\!\ed R^2}{R^2}+\!\ed\theta^2
		+\!\Lambda^2\!\pa{\!\ed\Phi+R\ed T}^2}\!,
\end{align}
where we have introduced $\Gamma(\theta)=\pa{1+\cos^2{\theta}}/2$ and $\Lambda(\theta)=2\sin{\theta}/\pa{1+\cos^2{\theta}}$.  This metric has two additional Killing fields relative to Kerr, making its isometry group $\mathsf{SO}(2,1)\times\mathsf{U}(1)$.  The important Killing field for our analysis is the \textit{dilation},
\begin{align}
\label{eq:Dilation}
	H_0=T\pd_T-R\pd_R,
\end{align}
whose finite form, $R\rightarrow cR$ and $T\rightarrow T/c$ for some constant $c$, is manifest in the metric \eqref{eq:NHEK}.\footnote{The other new Killing field is $H_-=\pa{T^2+1/R^2}\pd_T-2TR\pd_R-2/R\pd_\Phi$.  Together with $H_+=\pd_T$, the set $\cu{H_0,H_\pm}$ generates the conformal group $\mathsf{SO}(2,1)$, with commutation relations $\br{H_0,H_\pm}=\mp H_\pm$ and $\br{H_+,H_-}=2H_0$.}  The coordinates in Eq.~\eqref{eq:NHEK} cover the ``Poincar\'{e} patch'' $R>0$ of NHEK (region within the triangle in Fig.~\ref{fig:Collisions}).  The portions $R\rightarrow0,\,T\rightarrow\pm\infty$ are null surfaces called the future ($+$) and past ($-$) horizons.  The maximal analytic extension can be reached (e.g.) by the coordinate transformation (3.15) of Ref.~\cite{Lupsasca:2014}.

Consider a tensor field $W$ defined on extreme Kerr.  One may determine its near-horizon behavior by changing to scaling coordinates \eqref{eq:ScalingCoordinates} and expanding for small $\lambda$.  We will assume an expansion of the form
\begin{align}
\label{eq:Expansion}
	W&=\lambda^{-h}\pa{\bar{W}+\lambda\bar{W}^{(1)}+\frac{1}{2}\lambda^2\bar{W}^{(2)}+\dots},
\end{align}
where $h$ is some real number called the \textit{weight} of the field.  The leading piece $\bar{W}$ may be expressed as
\begin{align}
\label{eq:Limit}
	\bar{W}=\lim_{\lambda\rightarrow0}\lambda^{h}W,
\end{align}
with the limit at fixed scaling coordinates \eqref{eq:ScalingCoordinates}.  Since further rescalings $\lambda\rightarrow c\lambda$ do not change the limit, $\bar{W}$ must scale like $\bar{W}\rightarrow c^{-h}\bar{W}$ under $T\rightarrow T/c$ and  $R\rightarrow cR$, i.e.,\footnote{More generally, if $\phi_\sigma$ is a one-parameter group of diffeomorphisms with generating vector field $V$ \cite{Wald:1984} and $T$ is a smooth tensor field, then the limit $\bar{T}=\lim_{\sigma\rightarrow\infty}e^{-n\sigma}\phi_{-\sigma}^*T$ (if it exists) satisfies $\mathcal{L}_V\bar{T}=n\bar{T}$.  (That is, repeated application of a transformation promotes it to a symmetry.)  The diffeomorphism family \eqref{eq:ScalingCoordinates} has generator \eqref{eq:Dilation} and group parameter $\sigma=-\log\lambda$.}
\begin{align}
\label{eq:Self-similarity}
	\mathcal{L}_{H_0}\bar{W}=h\bar{W}.
\end{align}
Force-free solutions in NHEK obeying this condition were previously studied in Refs.~\cite{Lupsasca:2014,Zhang:2014,Lupsasca:2015,Compere:2016}; here, we show how it necessarily arises from the limit of smooth fields in Kerr.

The weight $h$ depends on the field being studied.  A field which is smooth on the future horizon of extreme Kerr can have weight no larger than its rank, since the Jacobian matrix \eqref{eq:Jacobian} relating the scaling coordinates \eqref{eq:ScalingCoordinates} to ingoing Kerr coordinates is $\mathcal{O}\!\pa{\lambda^{-1}}$.  Additional properties can reduce the maximum weight.  For example, the details of Eq.~\eqref{eq:Jacobian} show that two-forms (antisymmetric rank-2 tensors) have a maximum weight of 1 rather than 2.  The extreme Kerr metric has $h=0$, which is equivalent to the statement that the NHEK metric is dilation-invariant.  All NHEK fields that descend from regular Kerr fields in this way will be regular on the NHEK future horizon.

Fields with positive weight become null in the limit \eqref{eq:Limit}.  For example, consider a vector field $V$ of weight 1,
\begin{align}
	V=\lambda^{-1}\bar{V}+\mathcal{O}\!\pa{\lambda^0}.
\end{align}  
Since the metric has weight zero, the norm $V^2$ satisfies
\begin{align}
\label{eq:Null}
	V^2=\lambda^{-2}\bar{V}^2+\mathcal{O}\!\pa{\lambda^{-1}},
\end{align}
where $\bar{V}^2$ is constructed with the NHEK metric.  However, since $V^2$ is a regular scalar, its maximum weight is zero. It then follows from Eq.~\eqref{eq:Null} that $\bar{V}^2=0$.  Similarly, any scalar invariant of any $h>0$ field must vanish in the near-horizon limit.

Thus, smooth tensor fields $W$ on extreme Kerr have highly constrained near-horizon limits $\bar{W}$: following Eq.~\eqref{eq:Self-similarity}, they must be self-similar with some weight $h$, and if $h>0$, then they must also be null in the sense that all their scalar invariants vanish.  For perturbative calculations, it is useful to consider families $W(\lambda)$.  We define the weight and near-horizon limit in the same way as in Eq.~\eqref{eq:Limit}, but in general, there is no symmetry enhancement or tendency to become null.

The reader familiar with the study of critical phenomena in condensed matter systems or quantum field theory will recognize this discussion as the classification of infrared fixed points in terms of operator scaling dimensions. In the present astrophysical context, the infrared limit arises geometrically from the fact that we are scaling to a region of infinite redshift near the horizon.

\section{Maxwell fields}
\label{sec:Maxwell}

Consider a stationary, axisymmetric Maxwell field $F$ on extreme Kerr that is regular on the future horizon but not necessarily on the past horizon (or bifurcation two-sphere).  Generically, such a field has weight 1,
\begin{align}
	F=\lambda^{-1}\bar{F}+\mathcal{O}\!\pa{\lambda^0}.
\end{align}
The limiting field $\bar{F}$ must be stationary, axisymmetric, null, weight-1 self-similar $\pa{\mathcal{L}_{H_0}\bar{F}=\bar{F}}$, regular on the future horizon of NHEK, and closed $\pa{\!\ed\bar{F}=0}$.\footnote{We regard the vector potential as fundamental, so $\ed\bar{F}=0$ is just a consistency condition.}  These properties are highly constraining and reduce $\bar{F}$ to the form\footnote{An alternative derivation is to begin with the components of $F$ on the future horizon of extreme Kerr in a regular coordinate system and then take the near-horizon limit.}
\begin{align}
\label{eq:F_1}
	\bar{F}=A(\theta)\ed(T-1/R)\wedge\ed\theta.
\end{align}
The one-form $\ed(T-1/R)$ is tangent to the ingoing principal null congruence of NHEK (after raising the index).  The charge-current vector $\bar{J}$ is proportional to the null congruence,
\begin{align}
\label{eq:J_1}
	\bar{J}=-\frac{\pd_\theta\br{\Lambda(\theta)A(\theta)}}{2M^2\Gamma(\theta)\Lambda(\theta)}\ed(T-1/R).
\end{align}
(Here we give the lowered-index version as a one-form.)  Suppose instead that $F$ is weight zero, i.e., $A(\theta)=0$. Then
\begin{align}
\label{eq:Subleading}
	F=\hat{F}+\mathcal{O}(\lambda).
\end{align}
Again, $\hat{F}$ satisfies a long list of constraints: it must be stationary, axisymmetric, scale invariant $\pa{\mathcal{L}_{H_0}F=0}$, regular on the future horizon, and closed.  This again fixes the form up to a free function of the polar angle $\theta$,
\begin{align}
\label{eq:F_0}
	\hat{F}=B(\theta)\ed T\wedge\ed R+B'(\theta)(R\ed T+\ed\Phi)\wedge\ed\theta.
\end{align}
Note that this two-form is invariant under the full $\mathsf{SO}(2,1)\times\mathsf{U}(1)$ isometry group.  The one-form $R\ed T+\ed\Phi$ is proportional to the axial Killing field (after lowering the index), $\pd_\Phi=2M^2\Gamma(\theta)\Lambda(\theta)^2(R\ed T+\ed\Phi)$.  The charge-current is also proportional to the axial Killing field,
\begin{align}
\label{eq:J_0}
	\hat{J}=\frac{B'(\theta)\Lambda'(\theta)-B''(\theta)\Lambda(\theta)-B(\theta)\Lambda(\theta)^3}
		{\br{2M^2\Gamma(\theta)}^2\Lambda(\theta)^3}\frac{\pd}{\pd\Phi}.
\end{align}
In some contexts, one may also wish to demand regularity on the past as well as the future horizon.  It is easy to see that the $h=1$ fields in Eq.~\eqref{eq:F_1} all blow up on the past horizon.\footnote{This does not preclude their relevance to astrophysical black holes, which do not have a past horizon.}  Allowed fields are all of the $h=0$ ones as given in Eq.~\eqref{eq:F_0}.  Smoothness of these fields follows immediately from their full $\mathsf{SO}(2,1)\times \mathsf{U}(1)$ invariance and the fact that the $\mathsf{SO}(2,1)$ generator $H_+=\pd_T$ moves the location of the horizon.

The constraints of this section are purely kinematical.  In the next section, we impose further dynamical constraints in the two interesting cases of vacuum and force-free electrodynamics.

\subsection{Vacuum: Black hole Meissner effect}

Suppose the original two-form $F$ in extreme Kerr was generated by charge-current sources exterior to the black hole and not extending to the horizon.  Then the leading piece in the near-horizon limit must be a vacuum solution.  Setting the current \eqref{eq:J_1} to zero, we see that $A(\theta)=0$ and hence, $\bar{F}=0$ entirely.  So instead, we have Eq.~\eqref{eq:Subleading}, i.e., $h=0$.  Setting the current \eqref{eq:J_0} equal to zero gives an ordinary differential equation, whose general solution is
\begin{align}
	B(\theta)=-Q_E\cos{G(\theta)}+Q_M\sin{G(\theta)},
\end{align}
where
\begin{align}
	\cos{G(\theta)}=\frac{\Gamma(\theta)\Lambda(\theta)^2}{2}=\frac{1-\cos^2{\theta}}{1+\cos^2{\theta}}.
\end{align}
The constants $Q_E$ and $Q_M$ are the electric and magnetic charge associated with the solution (defined by the integral form of Gauss' law).  If these are nonzero, we would say that there is charge within the black hole, so we demand $Q_E=Q_M=0$, implying $\hat{F}=0$ entirely.  Thus, we have shown that the leading piece is $\mathcal{O}(\lambda)$, i.e., that in scaling coordinates, we have 
\begin{align}
	F=\mathcal{O}(\lambda).
\end{align}
In terms of ingoing coordinates $v,\tilde{r},\theta,\tilde{\phi}$ on Kerr (see Appendix~\ref{app:Regularity}), this means that
\begin{align}
	F_{v\tilde{\phi}}=F_{v\theta}&=F_{\theta\tilde{\phi}}=0,\\
	F_{\tilde{r}\tilde{\phi}}=2MF_{v\tilde{r}},\quad
	\pd_{\tilde{r}}F_{v\tilde{\phi}}&=0,\quad
	\pd_{\tilde{r}}F_{\theta\tilde{\phi}}=2M\pd_{\tilde{r}}F_{v\theta},
\end{align}
where everything is evaluated on the future horizon at $\tilde{r}=M$.  The first line is the statement that the pullback of $F$ to the future horizon vanishes, which we write as
\begin{align}
	F\big|_\mathcal{H}=0.
\end{align}
This is a covariant statement of the expulsion of magnetic flux.  In the classic derivation \cite{Bicak:1985} of the Meissner effect, one explicitly solves the vacuum Maxwell equation in the Kerr spacetime (at great effort) and finds that $F_{\theta\phi}=0$.  Here, we readily obtain a more general result by working with highly constrained fields in a much simpler metric.  See Refs.~\cite{Penna:2014,Bicak:2015,Hejda:2016} for other near-horizon analyses of the Meissner effect.

\subsection{Force-free}

If the magnetosphere is filled with tenuous plasma, then we want to impose the force-free equations, $\bar{F}\cdot\bar{J}=0$.  First, consider the generic case $h=1$.  Dotting Eq.~\eqref{eq:J_1} into Eq.~\eqref{eq:F_1} and recalling that $d(T-1/R)$ is null, we see that the leading two-form $\bar{F}$ is \textit{automatically force-free}.  This general type of null solution was studied in detail in Refs.~\cite{Brennan:2013,Brennan:2014,Gralla:2015a} and the NHEK solution \eqref{eq:F_1} was found explicitly in Ref.~\cite{Lupsasca:2015}.  Here, we reveal that solution to be the universal near-horizon limit for force-free plasma. 

Stationary, axisymmetric degenerate fields are normally represented by the flux function $\psi$, the field angular velocity $\Omega_F$, and the polar current $I$ (we use the conventions of Ref.~\cite{Gralla:2014}).  Taking the near-horizon limit of the general form for $F$ (Eq.~(65) of Ref.~\cite{Gralla:2014}) and using the regularity condition (Eq.~(104) of Ref.~\cite{Gralla:2014}) recovers the field in Eq.~\eqref{eq:F_1} and shows that the free function $A(\theta)$ is given by
\begin{align}
	A(\theta)=2M\br{\pd_\theta\psi(\Omega_F-\Omega_H)}\big|_{r=M}.
\end{align}
The values of these functions depend on the source of the external magnetic field.  Approximate analytic expressions for various field geometries (radial, parabolic, hyperbolic) are given in Ref.~\cite{Gralla:2015b}.

If $A(\theta)$ vanishes, then instead we have Eq.~\eqref{eq:J_0}, i.e., $h=0$.  Dotting Eq.~\eqref{eq:J_0} into Eq.~\eqref{eq:F_0}, we see that $\hat{F}$ is force-free when $B'(\theta)=0$, in which case it reduces to the simple form
\begin{align}
\label{eq:FFF}
	\hat{F}=-\frac{2}{\pi}Q_E\ed T\wedge\ed R.
\end{align}
This field has all the symmetries of NHEK and is electrically dominated, $\hat{F}^2<0$.  In flat Minkowski space, electrically dominated solutions are unstable to electric acceleration of charges, as seen by working in a local frame where the magnetic field vanishes.  However, this must be reconsidered in the presence of gravitational fields, in particular the strong fields of the NHEK geometry.  In principle, the charged particles in the plasma can be stabilized at points where the gravitational and electrical accelerations cancel.  Electrically dominated force-free electrodynamics also suffers from having nonhyperbolic field equations \cite{Komissarov:2002,Pfeiffer:2013}.  While this presents difficulties for making dynamical predictions, we see no reason why highly symmetric force-free fields such as the one in Eq.~\eqref{eq:FFF} could not appear as special solutions of a more general, dynamically well-posed theory.  Indeed, we have just seen that \textit{every} solution that is smooth and nonvanishing on all horizons must, in the near-horizon limit \eqref{eq:Limit}, reduce to the electrically dominated solution \eqref{eq:FFF}.

To explore this further, the Maxwell-Vlasov equations \cite{Choquet-Bruhat:2009} provide a physically reasonable kinetic description of the plasma and therefore seem likely to be suitably well-posed (see Ref.~\cite{Glassey:1968} for an existence result).  One could imagine finding a self-consistent particle distribution to accompany the electromagnetic field in Eq.~\eqref{eq:FFF} and then checking stability using the Maxwell-Vlasov equations.  We hope to tackle this problem in the future.

If one imposes $Q_E=0$, or otherwise discards the maximally symmetric solution, then $F$ has weight $h=-1$ instead.  This case was studied numerically in Ref.~\cite{Zhang:2014}.

\begin{figure*}
\centering
\subfigure[\ Near-horizon Penrose diagram]{\qquad\qquad\includegraphics[height=160pt]{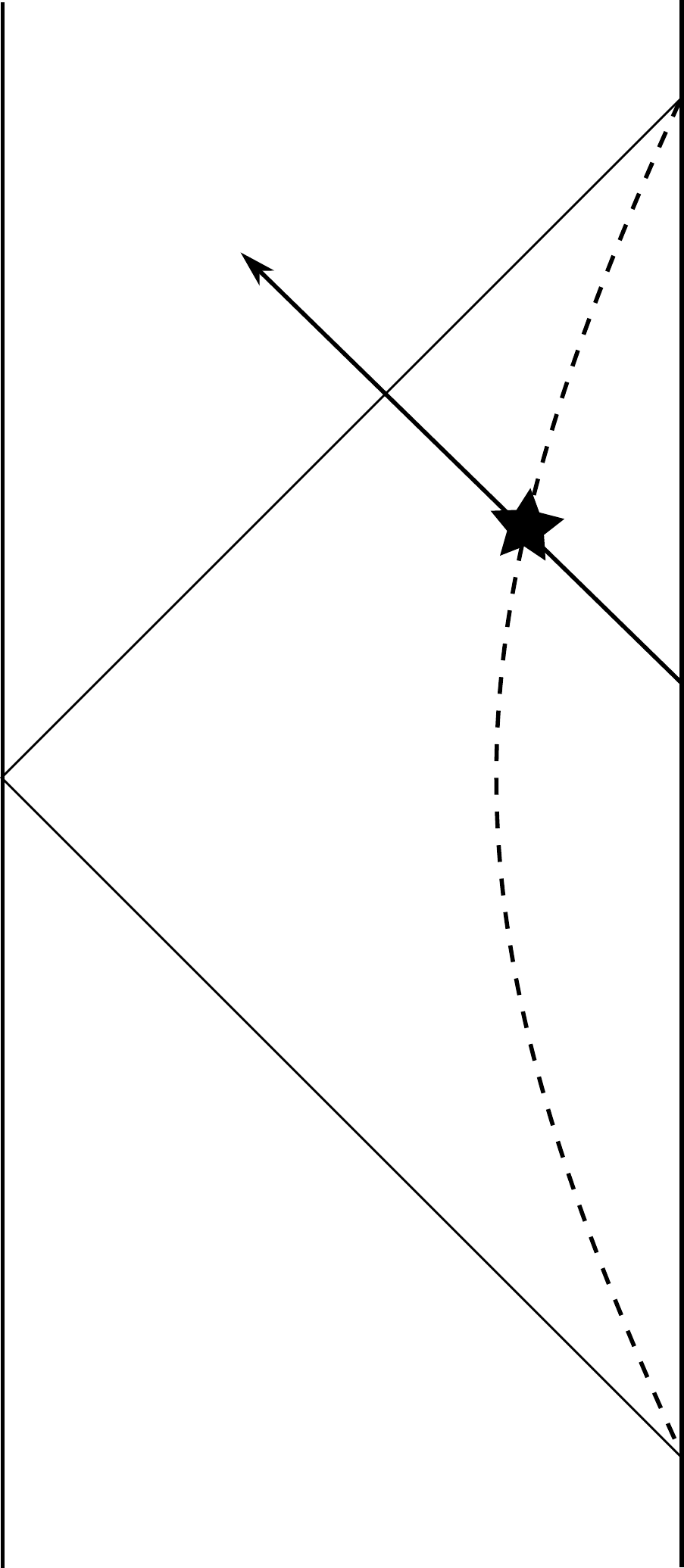}\qquad\qquad}
\qquad\qquad
\subfigure[\ Spatial diagram]{\includegraphics[height=80pt]{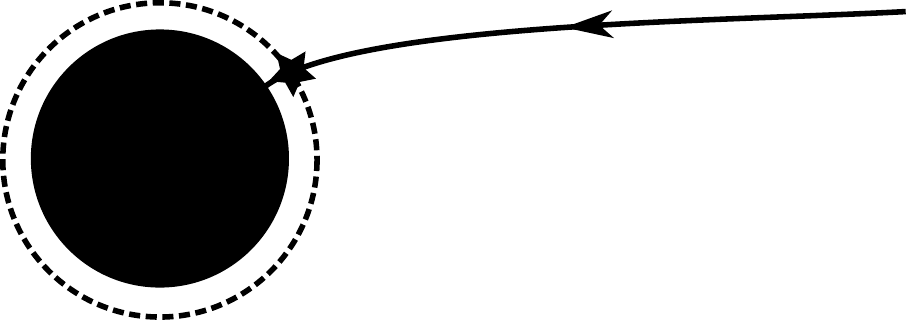}}
\caption{High-energy collisions near maximally spinning black holes.  The relative boost factor between tightly-bound orbits (dotted) and generic orbits (solid) can be arbitrarily large.  From a far-field point of view (right), the bound orbit is nearly null, while the plunging orbit is timelike.  In the near-horizon limit (left), the roles are reversed: the plunging orbit becomes null, while the bound orbit stays timelike.}
\label{fig:Collisions}
\end{figure*}

\section{Particle motion}
\label{sec:Particles}

Let $x^\mu(\tau)$ be a timelike worldline on extreme Kerr and denote the four-velocity $dx^\mu/d\tau$ by $u^\mu$.  If the worldline remains outside the horizon for all time, then it has no near-horizon limit, since $R\rightarrow\infty$ as $\lambda\rightarrow0$ at any fixed $r$.  The limit exists if the particle enters the black hole, but the worldline becomes null (and always a member of the ingoing principal null congruence).  To show this, we write $u$ in terms of its ingoing Kerr coordinates, change to the scaling coordinates \eqref{eq:ScalingCoordinates}, and expand.  This yields
\begin{align}
\label{eq:Plunge}
	u=\lambda^{-1}\bar{u}+\mathcal{O}\!\pa{\lambda^0},\quad
	\bar{u}=2M\Gamma(\theta)\pa{u\big|_H\cdot n}\!\ed(T-1/R),
\end{align}
where $u|_H$ is the four-velocity evaluated on the extreme Kerr horizon and $n=\ed\tilde{r}$ is the horizon normal.  The leading piece $\bar{u}$ is proportional\footnote{Since the horizon is a null surface and $u$ is timelike, $u|_H\cdot n$ is guaranteed to be nonzero for this plunging particle.} to $\ed(T-1/R)$, which we already noted points along the ingoing principal null congruence of NHEK.  This means that the worldline approaches the principal null congruence.\footnote{If we parametrize the worldline by $\tau/\lambda$ instead of proper time $\tau$, then the tangent vector smoothly approaches the null vector $\bar{u}$.}

The near-horizon limit is less constrained when one considers a family of trajectories on extreme Kerr.  For example, consider circular orbits in extreme Kerr, which have four-velocity
\begin{align}
	u=\frac{1}{e-\Omega\ell}(\!\pd_t +\Omega\pd_\phi),
\end{align} 
where $\Omega$ denotes the orbital frequency, while $e=-u\cdot\pd_t$ and $\ell=u\cdot\pd_\phi$ are the energy and angular momentum per unit mass.  Formulae for these quantities may be found in Ref.~\cite{Bardeen:1972}.  Work near (but outside) the horizon, $r=M(1+\delta)$ for $\delta>0$.  But as $\delta\rightarrow0$, we have $e\rightarrow\Omega\ell$ and the four-velocity blows up,
\begin{align}
\label{eq:FarOrbits}
	u=\frac{1}{\delta}\frac{4}{\sqrt{3}}\pa{\!\pd_t+\frac{1}{2M}\pd_\phi}
	+\frac{1}{6\sqrt{3}}\pa{4\pd_t-\frac{7}{M}\pd_\phi}+\mathcal{O}(\delta).
\end{align}

The blowup occurs because the family of timelike circular orbits of radius $r>M$ approaches a null circular orbit (the horizon generator) as $r\rightarrow M$.  We can work with a finite object by taking the near-horizon limit as follows.  Let $\delta=\lambda R$ so that $r=M(1+\lambda R)$, consistent with Eq.~\eqref{eq:ScalingCoordinates}.  Then change to scaling coordinates \eqref{eq:ScalingCoordinates} and expand in $\lambda$.  The result follows from Eqs.~\eqref{eq:KillingFields} and \eqref{eq:FarOrbits},
\begin{align}
\label{eq:NearOrbits}
	u=\bar{u}+\mathcal{O}(\lambda),\quad
	\bar{u}=\frac{1}{2M}\frac{4}{\sqrt{3}R}\pa{\!\pd_T-\frac{3}{4}R\pd_\Phi}.
\end{align} 
This is the four-velocity of a circular orbit of radius $R$ in NHEK \cite{Porfyriadis:2014}.

This connects to the fascinating observation that the center-of-mass energy of two finite-energy colliding particles in extreme Kerr can, for special initial conditions, be arbitrarily large \cite{Piran:1975,Banados:2009}.  The center-of-mass energy of two particles with four-momenta $p_1$ and $p_2$ is $\sqrt{-(p_1+p_2)^2}$.  Suppose that one particle is dropped in from infinity while the other particle orbits on a circular orbit of radius $r=M(1+\delta)$.  Then from Eq.~\eqref{eq:FarOrbits}, we see that the collision energy will scale as $1/\delta$ in the limit as $\delta\rightarrow0$, thereby growing arbitrarily large as the orbiting particle is tuned toward the horizon.

The effect may be understood from either a far-field or a near-horizon point of view (Fig.~\ref{fig:Collisions}).  From the far-field point of view, the infinite center-of-mass energy is blamed on the orbiting particle, which approaches a null orbit (the horizon) despite being massive.  On the other hand, in the near-horizon point of view, it is the \textit{infalling} particle that approaches a null geodesic (the principal null congruence of NHEK given in Eq.~\eqref{eq:Plunge}), while the orbiting particle occupies the harmless timelike circular orbit given in Eq.~\eqref{eq:NearOrbits}.  Neither picture is ``the correct one'', but the existence of the near-horizon limit seems intimately tied to the existence of the collisions in the first place.  One could have expected this behavior by reasoning that, since NHEK is a smooth spacetime at infinite boost relative to the Kerr exterior, it should be possible to have particles at infinite boost relative to particles from the exterior region.

\begin{figure*}
\centering
\subfigure[\ ISCO region Penrose diagram]{\qquad\qquad\label{fig:AGN-near}\includegraphics[height=200pt]{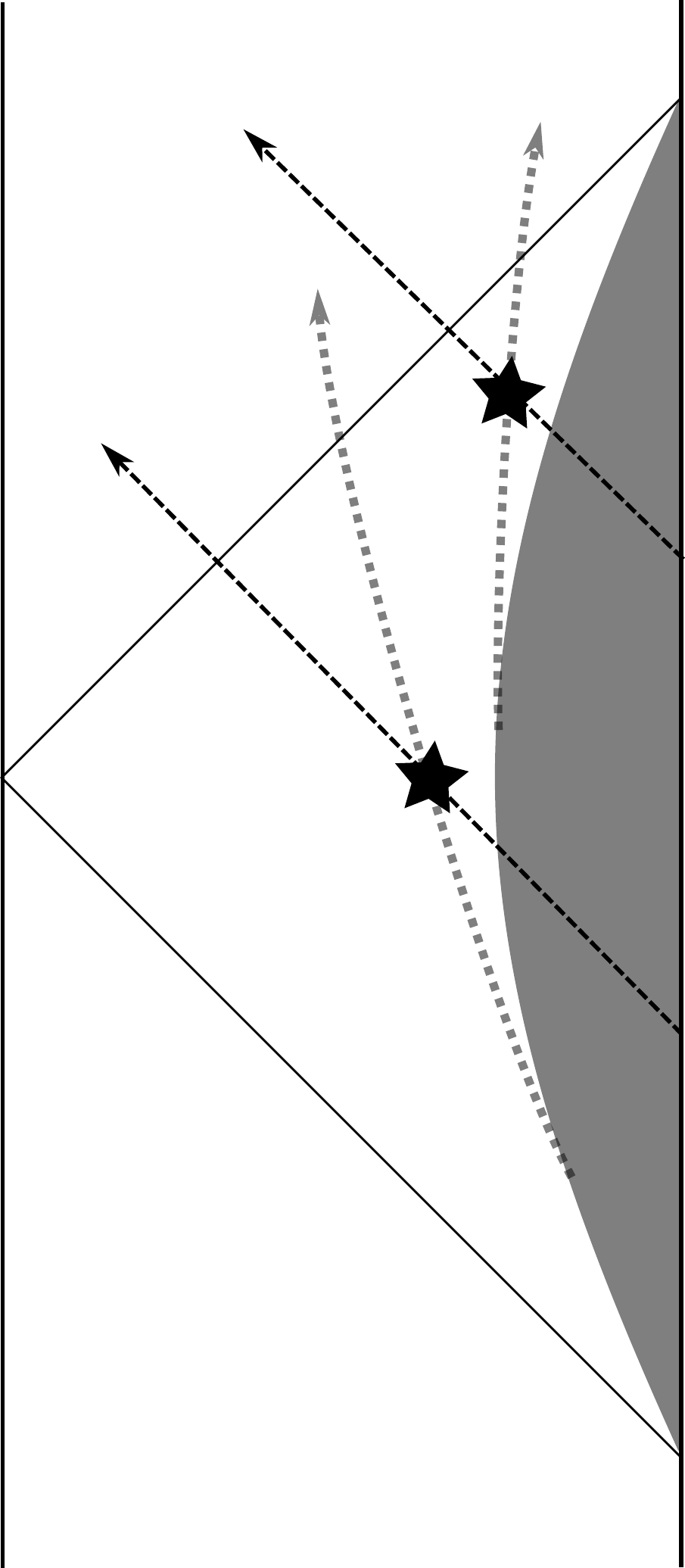}\qquad\qquad}
\qquad\qquad
\subfigure[\ Spatial diagram of an AGN]{\label{fig:AGN-far}\includegraphics[height=200pt]{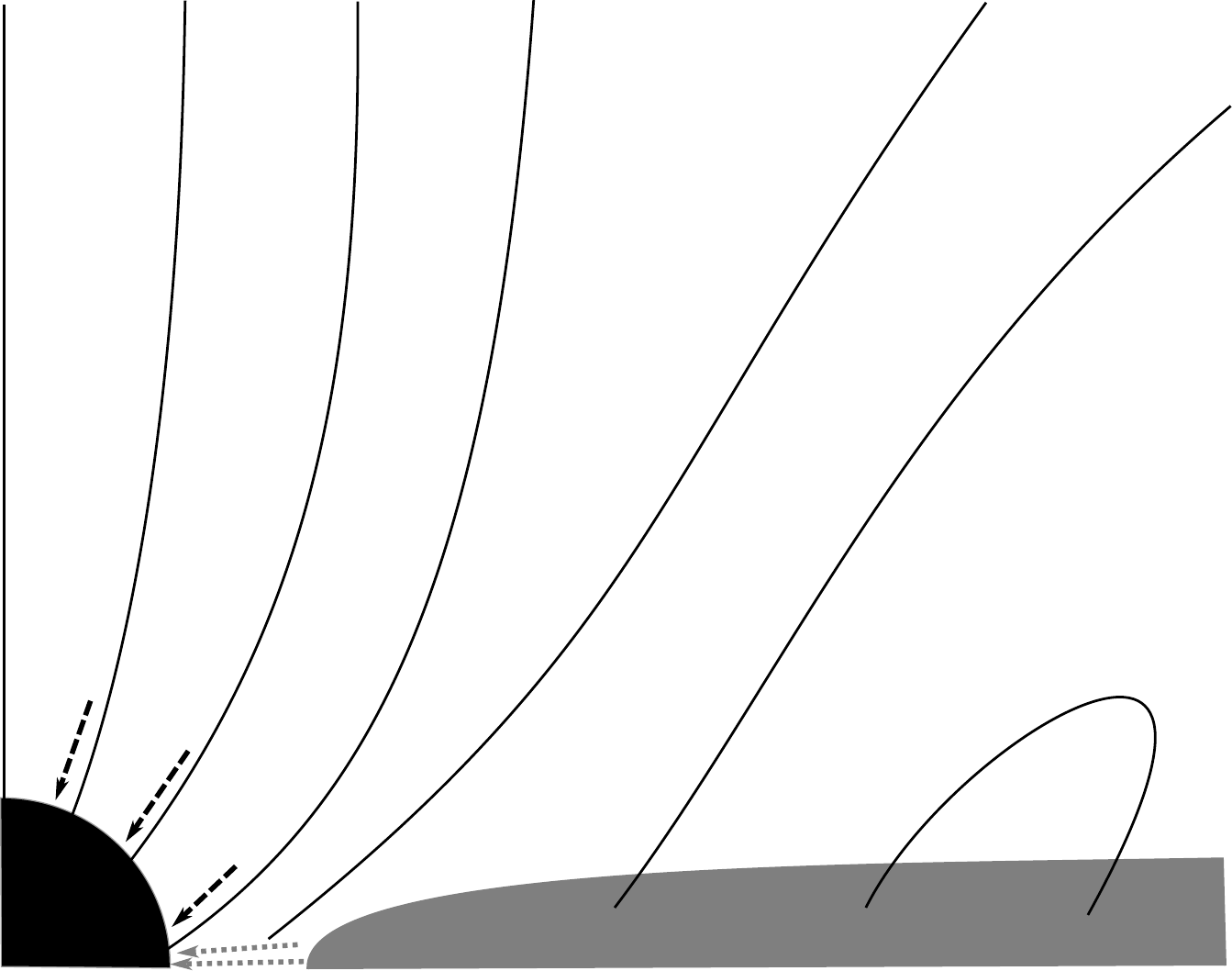}} 
\caption{In the BZ model of an AGN, dense matter from the disc falls in near the equator (dotted lines), while tenuous plasma flows in along field lines in the bulk (dashed lines).  The boost factor between these two flows is infinite in the extremal limit, suggesting the possibility of high-energy collisions.}
\label{fig:AGN}
\end{figure*}

\section{Near-extremal magnetosphere}
\label{sec:AGN}

Thus far, we have considered the near-horizon region of a precisely extremal black hole.  The near-horizon region of a near-extreme black hole is qualitatively different (described by the so-called near-NHEK metric \cite{Bredberg:2009}), but the region near the ISCO is described by the NHEK metric \eqref{eq:NHEK} (see Refs.~\cite{Hadar:2014,Gralla:2015c} and below).  This makes our conclusions generalize to the ISCO region of near-extreme black holes as follows.

Consider the Kerr metric in Boyer-Lindquist coordinates \eqref{eq:Kerr} and let
\begin{align}
\label{eq:Near-extreme}
	a=M\sqrt{1-(\kappa\lambda)^2}.
\end{align}
Here, $\kappa>0$ measures the black hole's deviation from extremality, and is held fixed as $\lambda\to0$.  We introduce new scaling coordinates [compare with Eq.~\eqref{eq:ScalingCoordinates}]:
\begin{align}
\label{eq:NewScalingCoordinates}
	T=\frac{\lambda^{2/3}t}{2M},\quad
	R=\frac{r-r_+}{\lambda^{2/3}r_+},\quad
	\Phi=\phi-\frac{t}{2M}.
\end{align}
Letting $\lambda\rightarrow0$ produces the NHEK metric \eqref{eq:NHEK}. While any choice $\lambda^p$ in \eqref{eq:NewScalingCoordinates} for $0<p<1$ would produce the NHEK metric, the power of $2/3$ in Eq.~\eqref{eq:NewScalingCoordinates} is designed so that the ISCO achieves a finite limit, approaching a timelike circular geodesic \cite{Hadar:2014,Gralla:2015c}.  This means that the edge of an accretion disk terminating at the ISCO will similarly survive the limit \eqref{eq:NewScalingCoordinates} and penetrate the NHEK region [Fig.~\ref{fig:AGN-near}].  Particles that spiral off the disk and fall into the black hole will also follow timelike orbits in NHEK before they exit the ISCO region by falling through $R=0$.

All of the analysis of Sec.~\ref{sec:Limits} goes through using Eqs.~\eqref{eq:Near-extreme} and \eqref{eq:NewScalingCoordinates} instead of $a=M$ and Eq.~\eqref{eq:ScalingCoordinates}.  More precisely, we consider smooth families of tensor fields, parametrized by the black hole spin $a\leq M$, each member of which is regular on the future horizon at $r=r_+=M+\sqrt{M^2-a^2}$.  (The ingoing coordinate components must be smooth functions of $v,\tilde{r},\theta,\tilde{\phi},$ and $a$.)  We then replace $\lambda$ by $\lambda^{2/3}$ in Eqs.~\eqref{eq:Expansion} and \eqref{eq:Limit}, take the ISCO limit using Eqs.~\eqref{eq:Near-extreme} and \eqref{eq:NewScalingCoordinates}, and conclude that the limiting fields are self-similar (Eq.~\eqref{eq:Self-similarity} still applies) and/or null according to their weight $h$.  Similarly, smooth families of timelike worldlines either become null or have no limit.  The ISCO itself is an example of a nonsmooth family (the $a\rightarrow M$ member is null) that tends to a timelike trajectory in the NHEK limit \eqref{eq:NewScalingCoordinates}.

The Blandford-Znajek model of an AGN features a magnetized accretion disk surrounded by a diffuse plasma [Fig.~\ref{fig:AGN-far}].  Near the equator, neutral matter from the disk falls into the black hole on geodesics that are slightly perturbed from the ISCO.  In the bulk region, charged particles move on magnetic field lines and continuously flow into the black hole \cite{Komissarov:2004}.  As previously mentioned, the infalling neutral matter is tuned to the ISCO region and follows a timelike trajectory in the NHEK limit obtained from Eq.~\eqref{eq:NewScalingCoordinates}.  On the other hand, the plasma particles are not tuned and limit to null trajectories.  Thus, the equatorial and bulk flows are at infinite relative boost in the extremal limit.  They are therefore at large relative boost near the ISCO of a near-extreme black hole, so any process that kicks a particle off of one flow and into the region of the other can potentially realize a high-energy collision.  Hence, AGN may naturally realize such collisions.

\acknowledgements{We are grateful to A.~Porfyriadis for useful conversations.  This work was supported in part by NSF grants 1205550 to Harvard University and 1506027 to the University of Arizona.}

\appendix

\section{Regularity}
\label{app:Regularity}

Ingoing Kerr coordinates $v,\tilde{r},\theta,\tilde{\phi}$ are related to Boyer-Lindquist coordinates by $v=t+r^*$, $\tilde{\phi}=t+r^\sharp$, and $\tilde{r}=r$ where for extreme Kerr $dr^*/dr=(r^2+a^2)/(r-M)^2$ and $dr^\sharp/dr=a/(r-M)^2$.  The Jacobian relating ingoing coordinates to scaling coordinates \eqref{eq:ScalingCoordinates} is given by
\begin{subequations}
\label{eq:Jacobian}
\begin{align}
	\ed v&=2M\br{\lambda^{-1}\ed\!\pa{T-\frac{1}{R}}+\frac{\ed R}{R}+\frac{\lambda}{2}\ed R}\\
	\ed\tilde{r}&=\lambda M\ed R\\
	\ed\tilde{\phi}&=\lambda^{-1}\ed\!\pa{T-\frac{1}{R}}+\ed\Phi.
\end{align}
\end{subequations}
Thus, the Jacobian matrix and its inverse are both $\mathcal{O}\!\pa{\lambda^{-1}}$, and transforming a rank-$N$ tensor from regular coordinates to scaling coordinates \eqref{eq:ScalingCoordinates} can introduce at most $N$ factors of $\lambda^{-1}$.

\section{Energy and angular momentum}

Near-horizon notions of energy and angular momentum become ``mixed up'' in a nontrivial way relative to far-field notions.  Here, we give a detailed treatment of this issue; see Ref.~\cite{Compere:2016} for a compatible treatment.

From Eq.~\eqref{eq:ScalingCoordinates}, we have
\begin{align}
\label{eq:KillingFields}
	\pd_\phi=\pd_\Phi,\quad
	k=\pd_t+\frac{1}{2M}\pd_\phi=\frac{\lambda}{2M}\pd_T,
\end{align}
where we introduce the horizon-generating Killing field $k$.  Near and far notions of angular momentum will therefore agree, while the near energy (conjugate to $H_+=\pd_T$) is in effect a renormalized corotating energy, infinitely deboosted to compensate for the fact that a true corotating observer would be moving at the speed of light.  The fact that this infinite deboost gives a good limit is intimately tied with the possibility of collisions of arbitrarily high energy when arbitrarily close to the horizon (Sec.~\ref{sec:Particles}).

The precise relationship between near and far notions of energy and angular momentum depends on the weight of the field under consideration.  Denote the stress tensor associated with the field $W$ by $T_{\alpha\beta}[W]$, which we suppose to be homogeneous and quadratic, $T_{\alpha\beta}[\lambda W]=\lambda^2T_{\alpha\beta}[W]$.  For stationary, axisymmetric solutions, the energy and angular momentum per unit time flowing out of the black hole are given by
\begin{align}
	\dot{\mathcal{E}}&=-\int{T^r}_t[W]\pa{r^2+a^2\cos^2{\theta}}\sin{\theta}\ed\theta\ed\phi,\\
	\dot{\mathcal{L}}&=\int{T^r}_\phi[W]\pa{r^2+a^2\cos^2{\theta}}\sin{\theta}\ed\theta\ed\phi,
\end{align}
evaluated at any radius $r$.  We define the near-horizon energy and angular momentum to be conjugate to $\pd_T$ and $\pd_\Phi$, respectively.  For stationary, axisymmetric fields, the fluxes relative to near-horizon time $T$ are then given by
\begin{align}
	\dot{\mathcal{E}}_N&=-4M^4\int{T^R}_T[\bar{W}] \Gamma(\theta)\sin{\theta}\ed\theta\ed\Phi,\\
	\dot{\mathcal{L}}_N&=4M^4\int{T^R}_\Phi[\bar{W}]\Gamma(\theta)\sin{\theta}\ed\theta\ed\Phi,
\end{align}
again at any radius $R$.  If $W$ has weight $h$ (i.e., Eq.~\eqref{eq:Limit} holds), then these are related by 
\begin{align}
	\dot{\mathcal{L}}&=\Omega_H\dot{\mathcal{L}}_N\lambda^{1-2h}\br{1+\mathcal{O}(\lambda)},\\
	\dot{\mathcal{E}}-\Omega_H\dot{\mathcal{L}}&=\Omega_H^2\dot{\mathcal{E}}_N\lambda^{2-2h}\br{1+\mathcal{O}(\lambda)},
\end{align}
recalling that $\Omega_H=1/(2M)$.  Note that $\dot{\mathcal{E}}$ denotes the energy flux per unit time $t$, while $\dot{\mathcal{E}}_N$ denotes the energy flux per unit near-horizon time $T$ (and likewise for the angular momentum fluxes).   The left-hand sides of these equations are numbers independent of $\lambda$, so we may take the $\lambda\rightarrow0$ limit on the right-hand sides.  In particular, the first equation shows that $\dot{\mathcal{L}}=0$ if $h<1/2$, whereas if $h=1/2$, then $\dot{\mathcal{L}}=\dot{\mathcal{L}}_N$.  If $h>1/2$, then $\dot{\mathcal{L}}_N$ must vanish and $\dot{\mathcal{L}}$ is not computable from the near-horizon limit of the field, requiring subleading corrections.  An analogous story holds for the ``corotating energy" flux $\dot{\mathcal{E}}-\Omega_H\dot{\mathcal{L}}$ at $h=1$.  In summary, the fluxes computable from the near-horizon limit of the field in each case are
\begin{align}
	h<1/2:&\quad\dot{\mathcal{L}}=\dot{\mathcal{E}}=0,\\
	h=1/2:&\quad\dot{\mathcal{E}}=\Omega_H\dot{\mathcal{L}}=\Omega_H\dot{\mathcal{L}}_N,\\
	1/2<h<1:&\quad\dot{\mathcal{E}}-\Omega_H\dot{\mathcal{L}}=0,\\
	h=1:&\quad\dot{\mathcal{E}}-\Omega_H\dot{\mathcal{L}}=\Omega_H^2\dot{\mathcal{E}}_N\leq 0,\label{eq:h=1}\\
	h>1:&\quad\textrm{nothing}.
\end{align}
In the second-to-last line we have noted that $\dot{\mathcal{E}}-\Omega_H\dot{\mathcal{L}}\leq0$, assuming the null energy condition (or equivalently, the first and second laws of thermodynamics).

In the text, we considered Maxwell fields with $h=1$ and $h=0$.  In the generic case $h=1$, the NHEK angular momentum flux always vanishes and the NHEK energy flux corresponds to the corotating flux $\dot{\mathcal{E}}-\Omega_H\dot{\mathcal{L}}$ via Eq.~\eqref{eq:h=1}.  The individual Kerr fluxes $\dot{\mathcal{E}}$ and $\dot{\mathcal{L}}$ are not computable from the near-horizon limit.  In the subleading case $h=0$, both fluxes vanish.

We now consider particles.  Let the four-velocity $u$ have weight $H$, i.e., $u=\lambda^{-H}\bar{u}\br{1+\mathcal{O}(\lambda)}$. Then to leading order in $\lambda$, the energy and angular momentum are given by
\begin{align}
	\ell=\lambda^{-H}\ell_N,\quad
	e-\Omega_H\ell=\frac{\lambda^{-H+1}}{2M}e_N,
\end{align}
where $\ell_N=\bar{u}\cdot\pd_\Phi$ and $e_N=-\bar{u}\cdot\pd_T$ (using the NHEK metric to contract) are the near-horizon angular momentum and energy, respectively.  Two interesting cases are
\begin{align}
	H=0:&\quad\ell=\ell_N,\quad e=\Omega_H\ell,\\
	H=1:&\quad e-\Omega_H\ell=e_N/(2M),\quad\ell_N=0.
\end{align} 
We note that there is an entirely different way of associating NHEK solutions with Kerr solutions, which is to apply a diffeomorphism relating NHEK to near-NHEK \cite{Bredberg:2009} and then interpret the fields as residing in the near-horizon region of a near-extreme black hole \cite{Hadar:2014,Zhang:2014,Lupsasca:2015,Hadar:2015}.  In this case, the relationships between near and far notions of energy and angular momentum are quite different.  This approach is less suitable for stationary solutions because the diffeomorphism involved does not preserve stationarity.\hfill\includegraphics[scale=0.03]{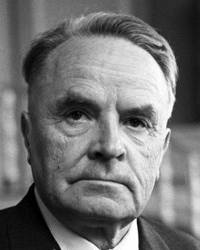}

\bibliographystyle{apsrev4-1}
\bibliography{nearmagnetosphere}

\end{document}